\documentclass[preprint,showpacs]{revtex4}
\usepackage[dvips]{graphicx}
\usepackage{amsmath,amsfonts,amssymb,bm,ulem}
\begin{document}

\title{Strong localization of positive charge in DNA induced by its interaction with environment}
\author{A. L. Burin and D. B. Uskov}
\affiliation{Department of Chemistry, Tulane University, New Orleans LA, 70118}
\date{\today}
\begin{abstract}
We investigate a quantum state of positive charge in DNA. A quantum state of electron hole is determined by the competition of the pi-stacking interaction $b$ sharing a charge between different base pairs and the interaction $\lambda$ with the local environment which attempts to  trap charge. To determine which interaction dominates we investigate charge quantum states in various $(GC)_{n}$ sequences choosing DNA parameters satisfying  experimental data for the balance of charge transfer rates $G^{+} \leftrightarrow G_{n}^{+}$, $n=2,3$. We show that experimental data can be consistent with theory  only assuming $b\ll \lambda$ meaning that charge is typically localized within the single $G$ site. Consequently as follows from our modeling consideration any DNA duplex including the one consisting of identical base pairs cannot be considered as a molecular conductor. Our theory can be verified experimentally measuring  balance of charge transfer reactions $G^{+} \leftrightarrow G_{n}^{+}$, $n \geq 4$ and comparing the experimental results with our predictions. 
\end{abstract}

\pacs{7080.Le, 72.20.Ee, 72.25.-b, 87.14.Gg}

\maketitle

\section{Introduction} 

Positive charge (hole) transfer in DNA is extensively investigated since its experimental discovery \cite{FirstExperiment}. This process can be responsible for the oxidative DNA damage \cite{FirstExperiment,Barton2,Barton3,GieseMB,Schuster1} and is possibly important for DNA repairing \cite{Raiskii,Taiwan}. Also an ability of DNA to promote long distant charge transfer can be used in molecular electronics applications \cite{MolecularWire1}. These studies raise the question, whether DNA is a molecular conductor. In molecular conductors (e. g. carbon nanotubes) charge is usually delocalized within several sites (monomers). Such system possesses a high conductivity similarly to metals. The alternative behavior takes place when the charge is typically localized within the single site and it hops to adjacent sites due to rare environment fluctuations. Based on the results of direct measurements of charge transfer in DNA \cite{FredMain} we argue in this paper that DNA is not a molecular conductor. 

DNA contains two different sorts of base pairs $AT$ and $GC$ forming quasi-random sequences. The lowest ionization potential is attributed to a $GC$ pair (essentially G-base \cite{Sugiyama}).  Since the electron transfer  integral $b$ between adjacent bases does not exceed $AT$ - $GC$ ionization potential difference $\Delta \sim 0.5$eV, the quantum state of charge in a static environment will be localized near some $G$ base and its localization length is comparable to the interbase distance \cite{Shapiro,local1}. Interaction with environment breaks down this localization, inducing charge hopping between quantum states localized at adjacent $G$ sites. This happens because of rare environment fluctuations supporting delocalization of charge between neighboring $G$ bases. Indeed, according to experimental studies \cite{GieseMB,Schuster1,FredMain} and theoretical models \cite{BixonJortner,BBR1} the sequence dependent charge transfer in DNA can be represented as the series of charge hops between adjacent $G$ bases serving as centers of localized states. An addition of $AT$ pair between $GC$ base pairs dramatically reduces the charge hopping rate \cite{GieseMB,BixonJortner,BBR1} and therefore the optimum base sequence for the efficient charge transfer consists of identical base pairs. We are going to study a charge (hole) quantum states in sequences of identical $GC$ pairs of various lengths, as most promising candidates for efficient charge transfer. 

The thermal energy at room temperature $k_{B}T \sim 0.026$eV is very small compared to at least one of the characteristic energies of the system including the reorganization energy $\lambda$ and the electron transfer integral $b$. The reorganization energy is caused by the electrostatic interaction of the hole with the environment and its minimum estimate is $\lambda \sim 0.25$eV \cite{Igor,abReview,LeBard,Voityuk,Berashevich}. The electron transfer integral $b$ is associated with the pi-stacking interaction of heterocyclic groups in DNA bases. It is defined as the gain in the hole energy due to its sharing between two adjacent DNA bases having equal energies with respect to the energy of the hole localized within a single $G$ base \cite{Voityuk}. The minimum estimate for the electron transfer integral between adjacent  $G$ bases is around $0.1$eV \cite{Voityuk}. Thus the thermal energy is, indeed, small compared to other characteristic energies and we  can use the ground state of the hole coupled to environment as the representative state. One should notice that the ground state can be used as a representative state, but it strongly differs from transition states responsible for the charge transfer. Since the hole transitions are quite rare the "typical" charge state can be described ignoring them.   

The spatial size of the hole ground state is determined by the competition of the delocalization of charge due to the pi-stacking interaction of heterocyclic groups belonging to adjacent bases and the localization caused by the environment polarization around the charge. The charge delocalization energy is characterized by the effective electron transfer integral $b$ and the localization energy is given by the environment reorganization energy $\lambda$ \cite{Igor,abReview,LeBard,Voityuk,Berashevich}. Delocalization of charge over $k$ base pairs leads to the gain in the kinetics energy $E_{del}\sim -2b+b/k^2$, similarly to the classical problem of particle in the box.  The reorganization energy scales with the size of charge wavefunction as $E_{loc}\sim -\lambda/k$ (see e. g. \cite{Conwell,Igor}). This dependence can be understood as following. The local environment polarization induced by the hole is proportional to the charge density $P \propto d \propto 1/k$, where $d$ stands for the local charge density. Consequently the interaction in each site having charge $e/k$ with the environment polarization is proportional to their product $pd \propto 1/k^2$ and the sum of all these $k$ interactions over all $k$ sites results in the dependence $\lambda/k$. This dependence takes place only in a one-dimensional system. Since at large $k$ the reorganization energy $-\lambda/k$ dominates over the kinetic energy $b/k^2$, the ground states is always localized and the localization radius can be estimated using the number of sites  $k$  minimizing  the  total energy 
\begin{equation}
E_{tot}(k) \approx -2b+\frac{b}{k^{2}} - \frac{\lambda}{k}. 
\label{eq:totEn}
\end{equation}
The minimum takes place at  $k=k_{*}\approx 2b/\lambda$ which is the number of DNA bases occupied by the hole in its ground state. At zero temperature the hole is localized, while at finite temperature it can hop between different states because of its interaction with the fluctuating environment. In the 
translationally invariant system ($(GC)_{n}$ or $(AT)_{n}$) the potential barrier separating two configurations can be estimated as the energy price 
for the environment fluctuation increasing the size of the wavefunction by one more site  compared to its optimum state $k_{*} \rightarrow k_{*}+1$,
which can be expressed as $E_{tot}(k_{*}+1)-E_{tot}(k_{*})$ Eq. (\ref{eq:totEn}). Such fluctuation creates the transition state and after the relaxation of environment the hole can arrive at the new equilibrium centered at different $G$ base. 
If $b\ll \lambda$ then this activation energy is given by approximately a quarter of the reorganization energy 
$\lambda/4 \gg k_{B}T$. Then we expect the hole mobility to obey the Arrhenius law as in insulators. In the opposite limit we obtain a very small value $\Delta\sim \lambda^{4}/(16b^{3})$,  which becomes negligible at moderately large $b/\lambda$. For instance if $b=1$eV the potential barrier for the hole transition becomes smaller than the thermal energy $k_{B}T \approx 0.026$eV already at $\lambda < 0.7eV$.  In this regime charge transport is weakly sensitive to the temperature as in metals. 

Therefore it is very important to determine the true relationship of $b$ and $\lambda$ in DNA. This relationship characterizes its conducting behavior. Indeed, in the regime $b > \lambda$ a DNA molecule made of identical base pairs would behave as a one-dimensional conductor, while in the opposite limit DNA cannot be treated as a molecular wire.  Existing estimates in literature \cite{Igor,abReview,LeBard,Voityuk,Conwell,Berashevich,Sugiyama} do not help much in determining  the relationship of $b$ and $\lambda$, because there is a large controversy between different studies. In particular, various estimates for the electron transfer integral $b$ range from $0.05$eV \cite{Voityuk} to $0.5$eV \cite{Sugiyama}.  Also all calculations of $b$ ignore vibrational rearrangements. Polar vibrational modes associated with the covalent bonds possess a high vibrational quantum energy $\hbar\omega \sim 0.3$eV $\gg k_{B}T$.  At room temperature these modes must remain in the ground state during charge tunneling. However since the tunneling of hole changes the equilibrium coordinates of polar vibrations they must tunnel together with the hole. This leads to the redefinition of the effective tunneling amplitude $b$ (if it is smaller then the reorganization energy $\lambda_{v}$) associated with the given vibrational mode \cite{classic1,comment1,ab_prl} as 
\begin{equation}
b_{eff} = b\cdot e^{-\frac{\lambda_{v}}{2\hbar\omega}}.
\label{eq:tunn_renorm}
\end{equation} 
To our knowledge this effect is ignored in all existing theoretical estimates of the electron transfer integral.

The calculations for the reorganization energy $\lambda$ associated with the classical interaction with solvent were made using the continuous medium approach. The estimates of $\lambda$ range from $0.25$eV \cite{Berashevich} to more than $1$eV \cite{Igor} due to uncertainty in the water dielectric constant value near the DNA molecule. It is hard therefore to select the right parameters based on existing theoretical results.

We suggest an alternative method to study the hole quantum state within the DNA molecule, using experimental data sensitive to the relationship of two key parameters of the theory $b$ and $\lambda$. Namely, we exploit the rate constants for the balance of charge transfer reactions between different $(GC)_{n}$ complexes, measured by Lewis and coworkers \cite{FredMain}
\begin{equation} \label{GrindEQ__5_} \begin{array}{l} {G^{+} +GG\mathop{\rightleftarrows }\limits_{k_{-t}^{GG} }^{k_{t}^{GG} } G+(GG)^{+}, \, \, \, \, \frac{k_{t}^{GG}}{k_{-t}^{GG}}=7.7\pm 1} \\ {G^{+} +GGG\mathop{\rightleftarrows }\limits_{k_{t}^{GGG} }^{k_{t}^{GGG} } G+(GGG)^{+}, \, \, \, \,\frac{k_{t}^{GGG}}{k_{-t}^{GGG}}=20\pm 1}. \end{array} \end{equation}
In the thermal equilibrium these ratios are determined by base pair  partition functions 
\begin{eqnarray}
r_{2}=\frac{k_{t}^{GG}}{k_{-t}^{GG}}=\frac{Z_{2+}Z_{1}}{Z_{2}Z_{1+}}, 
\nonumber\\ 
r_{3}=\frac{k_{t}^{GGG}}{k_{-t}^{GGG}}=\frac{Z_{3+}Z_{1}}{Z_{3}Z_{1+}}. 
\label{eq:rat1}
\end{eqnarray}
where $Z_{n+}$ stands for the partition function of $G_{n}$ sequence containing the single hole in it, while $Z_{n}$ is the partition function of the same base sequence, but without the hole.  Since $G$ and $G_{n}$ complexes are separated by some $AT$ bridge we can ignore their interactions.   

We assumed that the charge transfer between $G$ and $G_{n}$ complexes is very slow compared to the charge transfer rate inside the complex. This agrees with experiments showing the strong reduction of the charge transfer rate with the length of $AT$ separating $GC$ pairs \cite{GieseMB}. Consequently the contribution of highly excited states like the transition states for the charge transfer between complexes to the partition functions can be neglected so we can assume that the local equilibrium of charge within $G_{n}$ sequence is established. Accordingly one can use Eq. (\ref{eq:rat1}) for the ratio of charge transfer rates. 

Both ratios in Eq. (\ref{GrindEQ__5_}) are determined by parameters $b$ and $\lambda$ and the thermal energy at room temperature 
$k_{B}T \sim 0.026eV$. Below we calculate both ratios using tight binding model for $G_{n}$ complexes and standard linear response theory for  charge interaction with the environment \cite{abReview,Marcus}. Theory determines the domain of parameters $\lambda$ and $b$ satisfying experimental data Eq. (\ref{GrindEQ__5_}). {\it We demonstrate that any reasonable choice of $\lambda$ and $b$, satisfying Eq. (\ref{GrindEQ__5_}), corresponds to the regime $b \ll \lambda$, where the hole in its ground state is localized essentially in a single $G$ base (see Fig. \ref{fig:potential}). }

The manuscript is organized as following. In Sec. \ref{sec:model} we formulate and discuss the model for the hole in $G_{n}$ sequence coupled to the classical environment. In Sec. \ref{sec:expression} we derive expressions for rate ratios $r_{n}$.  In Sec. \ref{sec:part_func} we discuss the properties of the hole ground state in the domain of its parameters consistent with the experiment \cite{FredMain} and show that it is localized essentially within the single $G$ site. In Sec. \ref{sec:data_analysis} we discuss the partition functions of charge using the simple perturbation theory approach and predict the balance of charge transfer reactions between $G^{+}$ state and $G_{n}^{+}$ state for arbitrarily number $n$ of $G$ bases.  In Sec. \ref{sec:offdiag} we discuss the effect of correlations of environment polarizations on different $G$ sites on ratios $r_{n}$. In Sec. \ref{sec:Abases} we investigate the influence of $A$ bases surrounding $G_{n}$ sequences in the experiment \cite{FredMain} on our conclusions. In Sec. \ref{sec:conclusion} we resume our results and discuss the experiments, which can verify our theory and help to determine parameters $b$ and $\lambda$ more accurately. 

\section{Model}
\label{sec:model}

The chain of $n$ $GC$ base pairs can be described by the tight binding Hamiltonian coupled to the classical environment represented by coordinates $X_{i}$, $i=1, ... n$ for each DNA $G$ site 
\begin{eqnarray}
\widehat{H} = \widehat{H}_{hole} + \widehat{H}_{int},  
\nonumber\\
\widehat{H}_{hole}= -b\sum_{i=1}^{n-1}(c_{i}^{+}c_{i+1}+c_{i+1}^{+}c_{i}), 
\nonumber\\ 
\widehat{H}_{int} = \frac{1}{2\lambda}\sum_{i=1}^{n}X_{i}^{2}-\sum_{i=1}^{n}X_{i}n_{i},
\nonumber\\ 
n_{i}=c_{i}^{+}c_{i}. 
\label{eq:Hamiltonian}
\end{eqnarray}
Here $c_{i}$, $c_{i}^{+}$ are operators of creation and annihilation of electron hole in a site $i$. Classical coordinates $X_{i}$ describing the polar environment, 
including solvent and counterions, 
are directly coupled to the local charge density $n_{i}=c_{i}^{+}c_{i}$ and the parameter $\lambda$ stands for the reorganization energy associated with the interaction of charge and its environment. The environment energy is  expressed as a bilinear form with respect to solvent coordinates, which is justified by a standard assumption that polarization fields are small compared to atomic fields \cite{Marcus} so we ignore $X^{3}$ terms. Since the potentials $X$ include the contribution of counterions it is not clear whether the ion effect can be treated in the linear response approximation. However, we do not think that the contribution of ions can critically change our assumptions because as was demonstrated by Voityuk and coworkers \cite{voityuk4}  the contribution of ions does not exceed $20\%$ of the overall charge energy fluctuations, while the remaining $80\%$ of fluctuation are associated with water. As discussed below in Sec. \ref{sec:data_analysis} the validity of the linear response approximation does not crucially affect our consideration in the strong localization regime. 

We assume that only classical degrees of freedom with excitation energy comparable or less than the thermal energy are left in Eq. (\ref{eq:Hamiltonian}), while high energy modes are integrated out. This may lead to the renormalization of parameters in the system Hamiltonian Eq. (\ref{eq:Hamiltonian}) (see e. g. Eq. (\ref{eq:tunn_renorm}) and Refs. \cite{classic1,comment1,ab_prl}) and we assume that this renormalization is made. We do not include off-diagonal terms $X_{i}X_{j}$, $i\neq j$ into the Hamiltonian. This is somehow justified because they are smaller than the diagonal ones \cite{LeBard}. We will show below that for $G_{2}$ sequence the problem including off-diagonal terms can be reduced to the diagonal model Eq. (\ref{eq:Hamiltonian}) with the replacement of the single site reorganization energy $\lambda$ with the reorganization energy for charge transfer between adjacent sites. For $GGG$ sequence the similar replacement with removal off-diagonal terms remains a good approximation (see Sec. \ref{sec:offdiag}).  Therefore all our results are valid if we consider the more general interaction model with the replacement of the single site reorganization energy $\lambda$ with the charge transfer reorganization energy $\lambda_{*}$ as described below in Sec. \ref{sec:expression}. 

We assume the electron transfer integral $b$ to be independent of the environment fluctuations. Fluctuations of the electron transfer integral were treated as less significant compared to fluctuations in local site energies because the change of the site energy by more than the thermal energy strongly modifies the tunneling rate, while the change in the electron transfer integral requires the energy fluctuation comparable to the energy $\delta E \sim \hbar/\tau$ where $\tau$ is the tunneling time for the electron transition \cite{classic2}. This energy $\delta E$ can be comparable to the barrier height \cite{ab_7} which is much larger than the thermal energy. The above expectation conflicts with the molecular dynamics simulations of the electron transfer integral between hole states in adjacent G bases affected by the interaction with environment (see Ref. \cite{voityuk3} and references therein), where the remarkable effect of fluctuations  on the electron transfer integral has been found. However, molecular dynamics results should be considered with care because of the classical treatment of the nuclei motion, which can be partially of the quantum mechanical nature.  We estimated the effect of the distribution of the electron transfer integral $b$ on the ratios of charge transfer rates for the system described by Eq. (\ref{eq:Hamiltonian}). The distribution of $b$ was approximated by the Gaussian distribution with parameters $<b> \approx 0.046$eV and $\delta b \approx 0.064$eV \cite{voityuk3} while reorganization energy was taken as $\lambda=0.24eV$ \cite{voityuk4}.  The calculations result in the overestimated ratio of charge transfer rates $r_{2}$  Eq. (\ref{eq:rat1}) by the factor of five. This result queries  the accuracy of molecular dynamics approximation to fluctuations. Also the measurements of charge transfer efficiency between $G$ and $GGG$ sequences separated by $n$ $AT$ pairs \cite{giesenature}  can be successfully interpreted  assuming that the charge transfer integral is constant at least for $n<5$ $AT$ pairs \cite{ab_8}.

In the experiment  \cite{FredMain} $G_{n}$ sequence was surrounded by $A$ bases. The addition of $A$ bases surrounding $G_{n}$ sequences to our model Eq. (\ref{eq:Hamiltonian}) can change our results. The changes are associated with the possible sensitivity of the hole energy in the $G$ site to its neighbors and the ability of the hole to come virtually to the adjacent $A$ site. According to quantum chemistry calculations the ionization potential of the $G$ base depends on surrounding bases \cite{Voityuk1}. However, the  difference of ionization potentials for $GG^{+}G$, $AG^{+}G$ and $GG^{+}A$ does not exceed $0.1$eV. It is remarkably smaller than the reorganization energy and therefore we will ignore it. The error of semiempirical methods used to calculate the ionization potential is comparable to the maximum calculated effect $\sim 0.1$eV so we cannot consider the energy change  $0.1$eV or smaller. Screening of electrostatic interaction by water can also reduce the effect of neighbors on the $G^{+}$ state ionization potential.

The changes associated with the extension of the charge wavefunction to adjacent $A$ sites can be more important. However these changes do not affect our main conclusion about the strong localization of the charge ground state as shown in Sec. \ref{sec:Abases}. We ignore the second strand ($C_{n}$) because of the weak coupling between strands \cite{Voityuk} and the large difference of $G$ and $C$ ionization potentials exceeding $1$eV \cite{Sugiyama}. 

\section{Evaluation of rate ratios}
\label{sec:expression}

We study the ratios of charge transfer rates Eq. (\ref{eq:rat1}). Each partition function is given by $Z_{n}=\int dX_{1}...dX_{n}Tr e^{-\beta \widehat{H}_{n}}$, where $\widehat{H}_{n}$ is the Hamiltonian Eq. (\ref{eq:Hamiltonian}) describing $G_{n}$  sequence, trace is taken only over states with the single hole ($Z_{n+}$) or no holes ($Z_{n}$) and $\beta=1/(k_{B}T)$.  If there is no hole  calculations are reduced to multiple evaluation of a Gaussian integral leading to the expression
\begin{eqnarray}
Z_{n} = c^{n}, \, \, \, c=\sqrt{\frac{2\pi\lambda}{\beta}}.  
\label{eq:zero_part_trace}
\end{eqnarray}     
For the sequences containing a hole an analytical expression can be obtained only for $n=1$ 
\begin{eqnarray}
Z_{1+} = ce^{\beta\lambda/2}.  
\label{eq:one_part_trace1}
\end{eqnarray} 
For any number $n$ of $GC$ pairs calculations can be simplified integrating the partition function over  a ``center of mass'' coordinate $X_{1}+..X_{n}$, which is coupled to the conserving operator of the total number of particles $c_{1}^{+}c_{1}+..+c_{n}^{+}c_{n}=1$. In the case $n=2$ one can conveniently introduce the new coordinates as $X=X_{1}+X_{2}$ and $u=X_{1}-X_{2}$. Then the Hamiltonian of two base pairs (Eq. (\ref{eq:Hamiltonian}) for $n=2$) takes the form
\begin{equation}
\widehat{H}_{2} = \frac{X^{2}}{4\lambda} - \frac{X}{2} +\frac{u^{2}}{4\lambda} -\frac{u(n_{1}-n_{2})}{2} -b (c_{1}^{+}c_{2}+c_{2}^{+}c_{1}).   
\label{eq:Ham1}
\end{equation}
Eigenenergies of this Hamiltonian can be written as 
\begin{equation}
E_{\mp} = \frac{X^{2}}{4\lambda} - \frac{X}{2} + \frac{u^{2}}{4\lambda}  \mp \sqrt{\frac{u^2}{4}+b^2}.    
\label{eq:Ham1E}
\end{equation}
Accordingly the partition function takes the form 
\begin{eqnarray}
 Z_{2+} = \frac{1}{2}\int_{-\infty}^{+\infty}dXe^{\frac{-\beta X^{2}}{4\lambda} + \frac{\beta X}{2}}\int_{-\infty}^{+\infty}du e^{-\frac{\beta u^{2}}{4\lambda}}2\cosh\left(\beta\sqrt{\frac{u^{2}}{4}+b^{2}}\right), 
\label{eq:one_part_trace2before}
\end{eqnarray} 
The factor $1/2$ is concerned with the coordinate transformation. Performing integration over the center of mass coordinate $X$ we get   
\begin{eqnarray}
 Z_{2+} = (\sqrt{2}c)e^{\beta\lambda/4}\int_{-\infty}^{+\infty}du e^{-\frac{\beta u^{2}}{4\lambda}}\cosh\left(\beta\sqrt{\frac{u^{2}}{4}+b^{2}}\right). 
\label{eq:one_part_trace2}
\end{eqnarray} 
The ratio $r_{2}$ of reaction rates Eq. (\ref{eq:rat1}) takes the form 
\begin{eqnarray}
 r_{2} = \sqrt{\frac{\beta}{\pi\lambda}}e^{-\beta\lambda/4}\int_{-\infty}^{+\infty}du e^{-\frac{\beta u^{2}}{4\lambda}}\cosh\left(\beta\sqrt{\frac{u^{2}}{4}+b^{2}}\right). 
\label{eq:one_part_rat2}
\end{eqnarray} 

Consider $GG$ sequence with off-diagonal interaction of environment coordinates 
\begin{equation}
\widehat{H}_{env}= \frac{A_{0}(X_{1}^{2}+X_{2}^{2})}{2}+A_{1}X_{1}X_{2}.
\label{eq:off_diag_G_2H1}
\end{equation} 
The coefficients $\widehat{A}$ can be expressed in terms of the average coordinate fluctuation matrix $\widehat{B}$ as 
\begin{equation}
\left( \begin{array}{cc}
A_{0} & A_{1}  \\
A_{1} & A_{0} \end{array} \right) 
=  
\left( \begin{array}{cc}
B_{0} & B_{1}  \\
B_{1} & B_{0} \end{array} \right)
^{-1}, 
\label{eq:matrixA}
\end{equation}
where the elements of the matrix $\widehat{B}$ are defined as \cite{LeBard} 
\begin{equation}
k_{B}T B_{0}= <X_{1}^{2}>=<X_{2}^{2}>, \, \, k_{B}T B_{1}= <X_{1}X_{2}>.
\label{eq:matrixB}
\end{equation}
$<...>$ means the standard thermodynamics averaging with the Hamiltonian Eq. (\ref{eq:off_diag_G_2H1}).

Then we can still proceed to the new coordinates $X$ and $u$ defined as previously. The Hamiltonian can be written as  
\begin{equation}
\widehat{H}_{p1} = \frac{(A_{0}+A_{1})X^{2}}{4} - \frac{X}{2} +\frac{(A_{0}-A_{1})u^{2}}{4} -\frac{u(n_{1}-n_{2})}{2} -b (c_{1}^{+}c_{2}+c_{2}^{+}c_{1}).   
\label{eq:Ham11}
\end{equation}
Note that the Hamiltonian for the single site has different definition of  the diagonal term of the environment energy ($n =0, 1$ is the population of the single $G$ site)  
\begin{equation}
\widehat{H}_{s} = \frac{X^{2}}{2B_{0}} - nX. 
\label{Ham_soffd}
\end{equation}
This is because  the Hamiltonian must provide the right expressions for average squared polarizations in Eq. (\ref{eq:matrixB}). 
The partition functions involved in our consideration for $r_{2}$ can be expressed as
\begin{eqnarray}
Z_{1}= \sqrt{\frac{2\pi B_{0}}{\beta}}, \, \, 
Z_{1+}= \sqrt{\frac{2\pi B_{0}}{\beta}}e^{\beta B_{0}/2}, \, \,
Z_{2}=\frac{2\pi}{\beta\sqrt{A_{0}^{2}-A_{1}^{2}}},
\nonumber\\
Z_{2+} = \sqrt{\frac{\pi (B_{0}+B_{1})}{\beta}}e^{\beta (B_{0}+B_{1})/4}\int_{-\infty}^{+\infty}du e^{-\frac{\beta u^{2}}{4\lambda_{*}}}2\cosh\left(\beta\sqrt{\frac{u^{2}}{4}+b^{2}}\right),
\nonumber\\
\lambda_{*}=B_{0}-B_{1}.
\label{eq:G2offdpart}
\end{eqnarray} 
The expression for the ratio $r_{2}$ Eq. (\ref{eq:rat1}) takes the form 
 \begin{eqnarray}
 r_{2} = \sqrt{2\frac{\beta }{\pi (B_{0}-B_{1})}}e^{-\beta (B_{0}-B_{1}))/4}\int_{-\infty}^{+\infty}du e^{-\frac{\beta u^{2}}{4(B_{0}-B_{1})}}\cosh\left(\beta\sqrt{\frac{u^{2}}{4}+b^{2}}\right). 
\label{eq:one_part_rat2_offdiag}
\end{eqnarray} 
This expression has the form identical to Eq. (\ref{eq:one_part_rat2}) if the reorganization energy $\lambda$ is replaced with $\lambda_{*}=B_{0}-B_{1}$. The new parameter $\lambda_{*}$ is the reorganization energy for the charge transfer between adjacent sites. This is the relevant parameter which can be used also for $GGG$ sequence and more complicated sequences as it will be shown in Sec. \ref{sec:offdiag}. 
  
The expression for $r_{3}$ is more complicated. Below we give the result after integration over the coordinate of the center of mass $X=X_{1}+X_{2}+X_{3}$ and use two other coordinates $v=X_{1}-2X_{2}+X_{3}$ and $u=X_{1}-X_{3}$ in the case of the ``diagonal'' environment energy Eq. (\ref{eq:Hamiltonian})   
\begin{eqnarray}
r_{3}= \frac{\beta}{4\sqrt{3}\pi\lambda}e^{-\frac{\beta\lambda}{3}}
\int_{-\infty}^{+\infty}dv\int_{-\infty}^{+\infty}du e^{-\frac{\beta v^{2}}{12\lambda}-\frac{\beta u^{2}}{4\lambda} }
\cdot Tr\left(\exp(-\beta \widehat{V}_{3})\right), 
\nonumber\\
\widehat{V}_{3}=\left( \begin{array}{ccc}
\frac{v}{2}+\frac{u}{6} & -b & 0  \\
-b & -\frac{u}{3} & -b \\ 
0 & -b & -\frac{v}{2} +\frac{u}{6}\end{array} \right). 
\label{eq:G3integr}
\end{eqnarray}
Similarly one can obtain the expression for $r_{4}$ 
\begin{eqnarray}
r_{4}=\frac{1}{4}\left(\frac{\beta}{2\pi\lambda}\right)^{\frac{3}{2}}e^{-\frac{3\beta\lambda}{8}}\int_{-\infty}^{+\infty}dt\int_{-\infty}^{+\infty}dv\int_{-\infty}^{+\infty}du e^{-\frac{\beta t^{2}}{8\lambda}-\frac{\beta u^{2}}{4\lambda}-\frac{\beta v^{2}}{4\lambda}}
\cdot Tr\left(\exp(-\beta \widehat{V}_{4})\right);
\nonumber\\
\widehat{V}_{4}=\left( \begin{array}{cccc}
\frac{t}{4}+\frac{u}{2} & -b & 0 & 0  \\
-b & -\frac{t}{4}+ \frac{v}{2}& -b & 0 \\ 
0 & -b & -\frac{t}{4} -\frac{v}{2} & -b \\
0 & 0& -b & \frac{t}{4} -\frac{u}{2} 
\end{array} \right);
\nonumber\\
t=X_{1}-X_{2}-X_{3}+X_{4}, \, \, u=X_{1}-X_{4}, \, \, v=X_{2}-X_{3}.  
\label{eq:r4}
\end{eqnarray} 
Eqs. (\ref{eq:one_part_rat2}), (\ref{eq:G3integr}), (\ref{eq:r4}) were used for the numerical evaluation of partition functions and ratios $r_{2}$, $r_{3}$ and $r_{4}$.

\section{Definition of parameters consistent with experiment} 
\label{sec:data_analysis1}

\subsection{Numerical results}
\label{sec:part_func}
\begin{figure}[b]
\centering
\includegraphics[width=15cm]{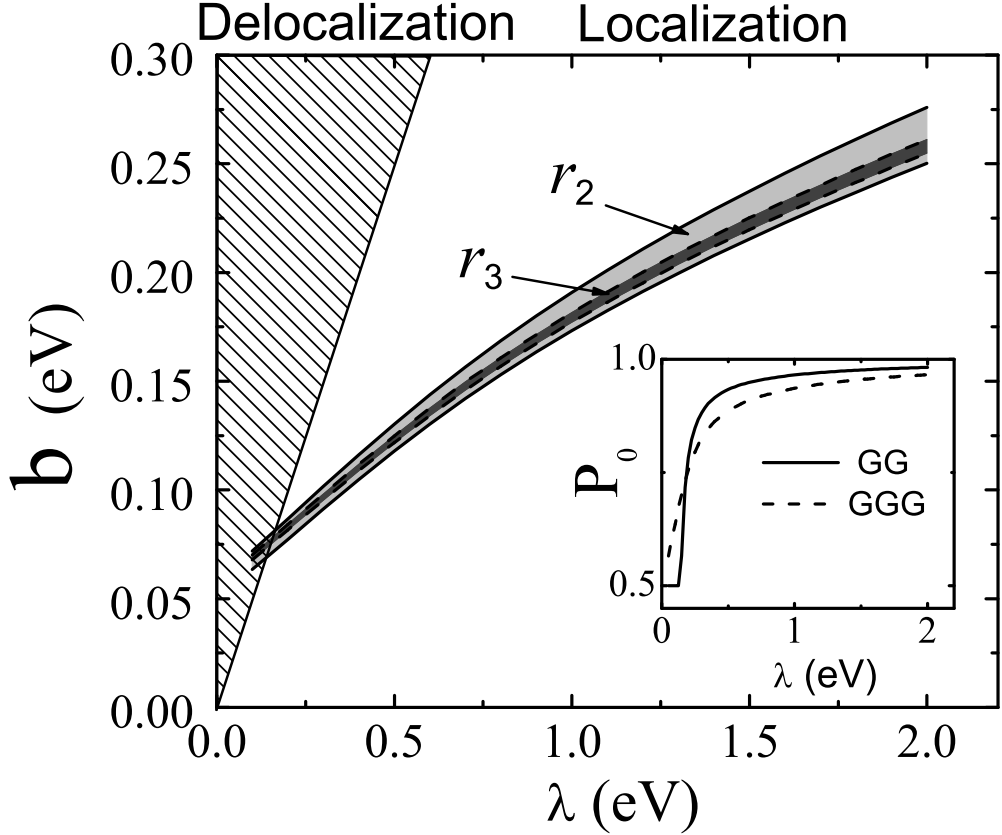}
\caption{The domains consistent with the experimental ratios of reaction rates Eq. (\ref{GrindEQ__5_}), dark grey for $GGG$ and  light grey for $GG$. Inset shows the fraction of the quantum charge state belonging to the central site ($P_{0}$) vs. the reorganization energy $\lambda$. Charge transfer integral $b$ is determined in the way to satisfy Eq. (\ref{GrindEQ__5_}). We assumed that the system is nearly in its ground state.  
\label{fig:potential} }
\end{figure}

We have performed numerical evaluations of ratios in Eq. (\ref{eq:rat1}) to find domains of parameters $b$ and $\lambda$ satisfying Eq. (\ref{GrindEQ__5_}) and show these domains in Fig. \ref{fig:potential}. The upper (lower) border of each domain is defined by the maximum (minimum) value of ratios $r_{2}$ and $r_{3}$ Eq. (\ref{eq:rat1}) within the experimental error ($8.7$ ($6.7$) for $GG$ and $21$ ($19$) for $GGG$). The acceptable domain of parameters for a $GGG$ sequence fully belongs to the corresponding domain for a $GG$ sequence. Thus the domains for $GG$ and $GGG$ base sequences are completely consistent with each other. Therefore we cannot determine parameters $\lambda$ and $b$ better then using the ``dark'' domain for $GGG$. This information is still sufficient to consider the localization of the hole wavefunction in $G_{n}$ aggregates.
  
Since the thermal energy $k_{B}T \approx 0.026$eV is smaller than at least one of two other characteristic energies of the system (remember that the minimum estimate for the reorganization energy is $\lambda \sim 0.25$eV \cite{Berashevich}) we can characterize the wavefunction using the system ground state at coordinates $X_{i}$ ($i=1, ... n$) minimizing the ground state energy. In the relevant domain of parameters in Fig. \ref{fig:potential} ($\lambda > 0.25$eV) the ground state wavefunction is centered at one of $G$ bases (left or right ones for a $GG$ sequence and the central one for a $GGG$ sequence). We describe this state by the probability $P_{0}$ for the particle in the ground state to be in this central site. This probability can be calculated using the relationship   
\begin{equation}
P_{0}=\frac{X_{i}}{\lambda}, 
\label{eq:coord_reorg_en}
\end{equation}
where $i$ is the central site. This equation can be derived as following. 
Coordinates $X_{j}$ ($j=1,...n$) in the energy minimum satisfy the condition 
\begin{eqnarray}
\frac{\partial \left<g\left|\widehat{H}\right|g\right>}{\partial X_{j}}= 0, 
\label{eq:energymin1}
\end{eqnarray}
where $|g>$ is the wavefunction of the hole ground state at given coordinates $X_{j}$. 
Using Eq. (\ref{eq:Hamiltonian}) one can rewrite Eq. (\ref{eq:energymin1}) as
\begin{eqnarray}
0=\frac{X_{j}}{\lambda} - <n_{j}>. 
\label{eq:wnwrgymin2}
\end{eqnarray}
The average population of the central site $i$ ($<n_{i}>$) is equal to the probability $P_{0}$ to find the hole there. Accordingly we end up with Eq.(\ref{eq:coord_reorg_en}). 
  
Consider the ground state wavefunction for the hole within the $GG$ sequence.  The expression for the hole ground state energy of the $GG$ sequence at arbitrary coordinates $X_{1}$, $X_{2}$ reads  
$E_{2}=\frac{X_{1}^{2}+X_{2}^{2}}{2\lambda}-\frac{X_{1}+X_{2}}{2}-\sqrt{\frac{(X_{1}-X_{2})^2}{4}+b^2}$.
In the regime of interest $2b<\lambda$ (see Fig. \ref{fig:potential}) the minimum of energy is given by 
\begin{equation}
E_{2min}=-\frac{\lambda}{2}-\frac{b^2}{\lambda},   
\label{eq:pairenergymin}
\end{equation}
and it is realized at $X_{1}=\lambda/2\pm\sqrt{(\lambda/2)^2-b^2} = \lambda-X_{2}$. Accordingly 
\begin{equation}
P_{0}=\frac{X_{1}}{\lambda}=\frac{\lambda+\sqrt{\lambda^2-4b^2}}{2\lambda}. 
\label{eq:p02}
\end{equation}
Note that if $2\lambda<b$ the ground state wavefunction is symmetric in the minimum ($X_{1}=X_{2}=\lambda/2$) and the energy of this state is given by  
\begin{equation}
E_{2symm}=-\frac{\lambda}{4}-b.  
\label{eq:pairenergyminsaddle}
\end{equation}
Consequently $P_{0}=1/2$ (see Fig. \ref{fig:potential}).  
In the case of $2b<\lambda$ this symmetric state is the transition state for the charge transfer $G^{+}G \leftarrow\rightarrow GG^{+}$ (saddle point in the energy function $E_{2}(X_{1},X_{2})$ between the energy minima centered in the first and the second $G$ bases). For $GGG$ complex the probability $P_{0}$ that the hole  resides in the central site has been evaluated numerically. 

Both probabilities $P_{0}$ obtained for the $\lambda$ - $b$ line corresponding to the ratio $r_{2}=7.7$ Eq. (\ref{eq:rat1}) are shown in the inset in Fig. \ref{fig:potential}. It is clear from this graph that for both $GG$ and $GGG$ sequences the wavefunction of the hole is essentially localized within the single $G$ site. For instance at the minimum value of $\lambda \sim 0.25eV$ we have $85\%$ and $78\%$ of the probability to find the particle in that site for $GG$ and $GGG$ sequences, respectively.  These probabilities increase for $\lambda=1$eV to $96\%$ and $94\%$, respectively. 
Thus we come to the important conclusion that the wavefunction of hole is essentially localized in the single $G$ site for $G_{n}$ sequences. This conclusion differs from the predictions of previous work \cite{Berashevich,Conwell,Taiwan} where the the quantum state of the hole was represented as a polaron of an intermediate range shared between several base pairs. 

\begin{figure}[b]
\centering
\includegraphics[width=15cm]{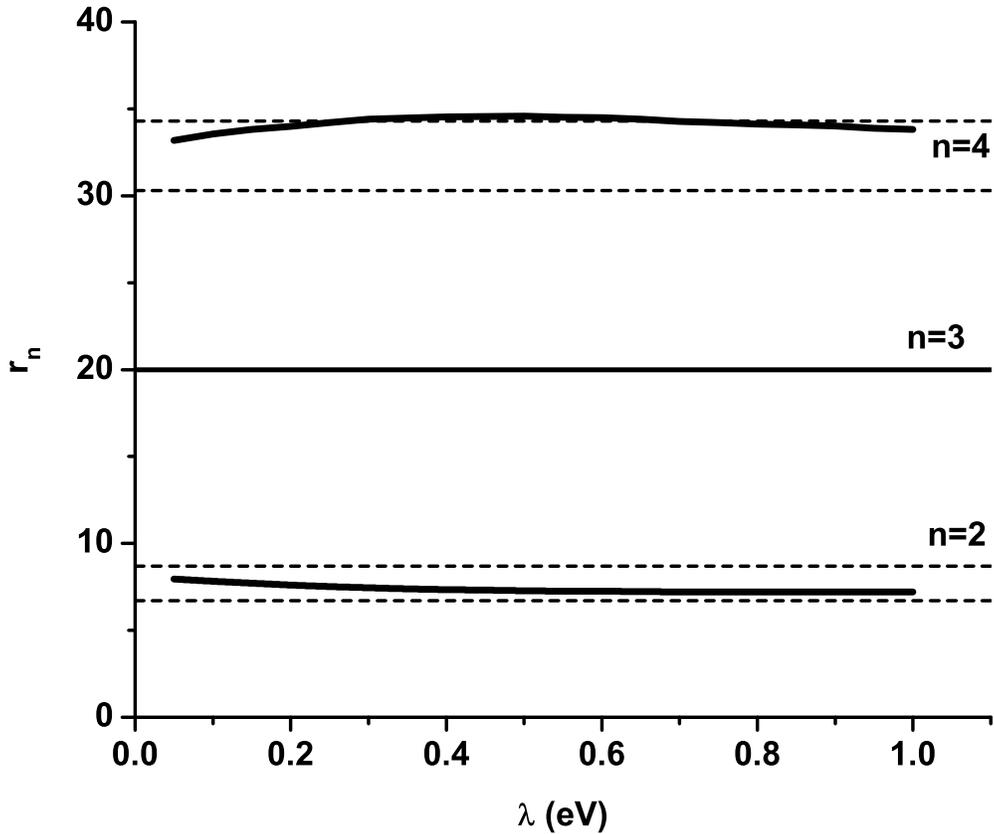}
\caption{Predictions for the ratio $r_{4}$ using parameters $b$ and $\lambda$ related by the experimental condition $r_{3}(b, \lambda)=20$ Eq. (\ref{GrindEQ__5_}). Solid line denoted by $n=4$ shows the predicted value of $r_{4}$ while two dashed lines show upper and lower boundaries for $r_{4}$ defined in accordance with Eq. (\ref{eq:arithm_ser1}). Results for $n=2$ calculated under the same assumptions are shown for comparison. Upper and lower boundaries for $r_{2}$ are given in accordance with the experiment \cite{FredMain}, Eq. (\ref{GrindEQ__5_}).  
\label{fig:G4} }
\end{figure}

In addition to the analysis of the $G_{2}^{+}$ and $G_{3}^{+}$ complexes we have performed the calculations for $G_{4}^{+}$ complexes for experimental verification of our theory. The partition function for $G_{4}^{+}$ complex can be evaluated integrating first the center of mass position similarly to our previous calculations. The predictions for the ratio $r_{4}$ are shown in Fig. \ref{fig:G4} by the solid line denoted as $n=4$. The parameters $\lambda$ and $b$ for calculations were obtained using the experimental condition  $r_{3}(\lambda, b)=20$ Eq. (\ref{GrindEQ__5_}). Using these parameters we calculated the ratio $r_{2}$ (solid line marked by $n=2$) and ratio $r_{4}$ (solid line marked by $n=4$) Eq. (\ref{eq:r4}).

According to Fig. \ref{fig:G4} the ratio $r_{4}$ does not change much in the whole domain of parameters $\lambda$ and $b$ satisfying the balance equations Eq. (\ref{GrindEQ__5_}). We can estimate it to be around $33$ and this is our prediction for the further experimental verification.

\subsection{Understanding of the balance of charge transfer reactions}
\label{sec:data_analysis}

\begin{figure}[b]
\centering
\includegraphics[width=15cm]{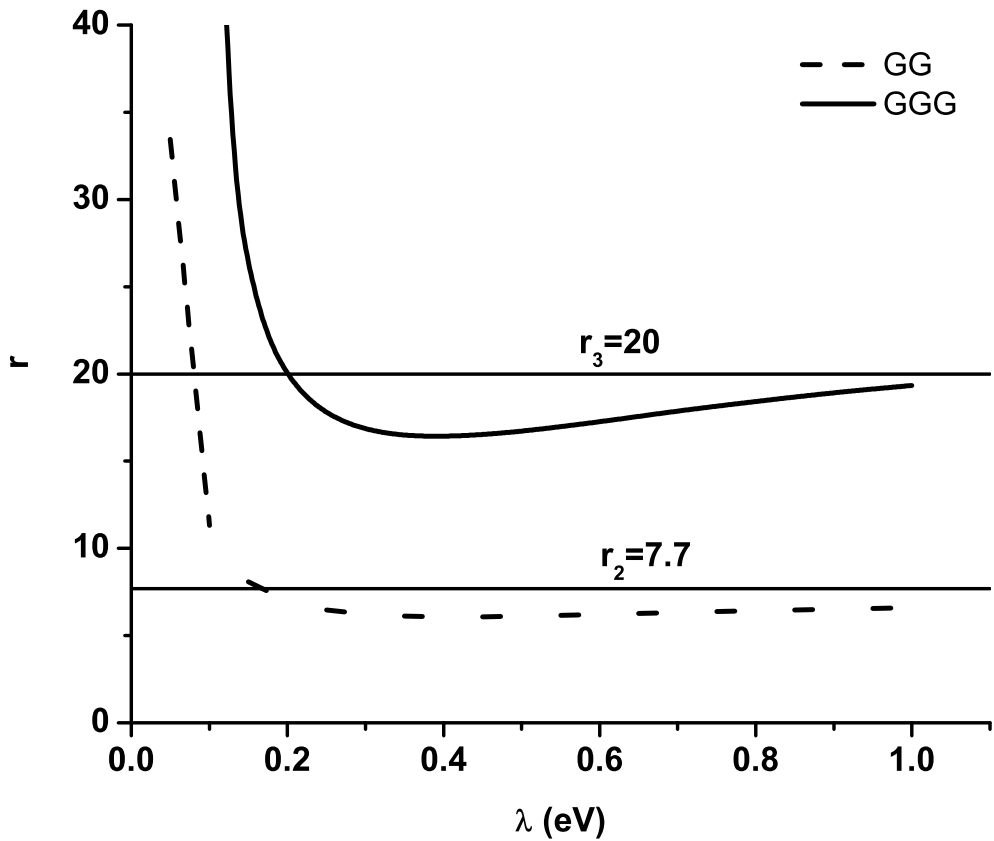}
\caption{Comparison of perturbation theory approaches with numerical simulations for ratios $r_{1}$ and $r_{2}$. Curves show predictions of Eq. (\ref
{eq:theorGGG}) for parameters $\lambda$ and $b$ chosen to satisfy Eqs. (\ref{eq:one_part_rat2}), (\ref{eq:G3integr}), respectively  ($r_{2}(\lambda, b)=7.7$, $r_{3}(\lambda, b)=20$).  
\label{fig:one_part_rat2} }
\end{figure}

An impressive consistency between $GG$ and $GGG$ base sequences in Fig. \ref{fig:potential} is not accidental coincidence and can be explained by the strong localization of charge wavefunctions. In the regime of strong localization the partition function $Z_{n+}$ for $n\geq 2$ can be represented as the sum of $n$ contributions of energy minima corresponding to wavefunctions centered in all $n$ $G$ sites with coordinates $X$ realizing the corresponding energy 
minimum $X_{i} \approx \lambda \gg X_{k}$, $k\neq i$ for the state centered at site $i$. Since in the zeroth order approximation in $b/\lambda$ each quantum state is localized at one site we can neglect the difference in preexponential factors for the case of $b=0$ and approximate the partial $i^{th}$ contribution to the partition function as $Z_{n}^{i}=c^{n}e^{-\beta E_{i}^{(0)}}$ (see Eq. (\ref{eq:zero_part_trace})), where $E_{i}^{(0)}=-\lambda/2$ is the energy of the ground state for coordinates $X$ realizing the local minimum in the absence of electronic coupling ($b=0$). Second order correction in $b$ to the energy $E_{i}^{(0)} = -\lambda/2$, which is the first non-vanishing correction, is important because it appears in the exponent $e^{-\beta E}$ and it is multiplied by the large factor $\beta$. For two states at the edges this correction adds to the hole energy as $E_{1}^{(1)} = E_{n}^{(1)} \approx -\lambda/2-b^{2}/\lambda$, which coincides with the ground state energy for $GG$ Eq. (\ref{eq:pairenergymin}). For $n-2$ remaining states the correction to the energy should be doubled because of the addition of contributions of two neighbors so we got $E_{i}^{(1)} = -\lambda/2 - 2b^{2}/\lambda$, $1<i<n$. The change in equilibrium environment coordinates can be neglected because near the energy minimum it leads to the effect of order of $(b/\lambda)^{4}$. 
Consequently, we can approximate the ratio $r_{n}$ (cf. Eq. (\ref{eq:rat1})) as 
\begin{eqnarray}
r_{n}=\frac{Z_{n+}Z_{1}}{Z_{n}Z_{1+}} \approx 2e^{\beta b^2/\lambda}+(n-2)e^{2\beta b^2/\lambda}. 
\label{eq:theorGGG} 
\end{eqnarray}
These predictions are compared with the numerical calculations in Fig. \ref{fig:one_part_rat2}. One can conclude that perturbation theory works reasonably well in the parameter domain of interest corresponding to $\lambda>0.25$eV for $n=2$ and $n=3$. 

In particularly, it follows from Eq. (\ref{eq:theorGGG}) that  $(Z_{3+}Z_{1}/(Z_{3}Z_{1+}) \approx (Z_{2+}Z_{1}/(Z_{2}Z_{1+})+((Z_{2+}Z_{1}/(2Z_{2}Z_{1+}))^2)$. Indeed, this relationship is satisfied for the experimental values of ratios within the accuracy of the experiment. This explains the consistency of domains for $GG$ and $GGG$ Fig. \ref{fig:potential}. Using Eq. (\ref{eq:theorGGG}) one can predict that ratios $r_{n}$ form arithmetic series, which can be expressed as 
\begin{equation}
r_{n}=7.7+12.3\cdot (n-2). 
\label{eq:arithm_ser1}
\end{equation}
For balance between $G$ and $GGGG$ sequence we predict the ratio $r_{4}=2r_{3}-r_{2}=32.3\pm 2$. This estimate agrees with our numerical calculations for the $(GGGG)^{+}$ partition function and reaction rate ratio $r_{4}$ (see Fig. \ref{fig:G4}). 

The perturbation theory analysis described above is approximately valid even if the energy of solvent fluctuations Eq. (\ref{eq:Ham1}) can not be expressed as the bilinear function of coordinates $X$. One can still define the reorganization energy $\lambda$ as the energy difference of adjacent $G$ bases in $G^{+}G$ sequence induced by the environment polarization around the charge localized at the only one (left) base. Then all our conclusions about localization remain valid, but the potential barriers for charge transfer will be larger than in the case of the "harmonic" potential ($\lambda/4$) because higher order nonlinear terms make the barrier sharper. 

Since  perturbation theory works reasonably well in the parameter domain of interest, we will use it to study more complicated questions of the charge transfer reaction balance in the presence of off-diagonal interaction of solvent polarizations (cf. Eq. (\ref{eq:off_diag_G_2H1})) and in the presence of $A$ bases surrounding $G_{n}$ complexes in experiments \cite{FredMain}. This analysis is reported below in Secs. \ref{sec:offdiag}, \ref{sec:Abases}, respectively.  

\subsection{Perturbation theory in case of off-diagonal correlations of environment polarizations}
\label{sec:offdiag}

The most general Hamiltonian of environment in the bilinear approach can be expressed as 
\begin{equation}
\widehat{H}_{env} = \frac{1}{2}\sum_{i,j=1}^{n}A_{ij}X_{i}X_{j},
\label{eq:H_gen_offdiag}
\end{equation}
where the interaction matrix $\widehat{A}$ is defined as the inverse correlation matrix $\widehat{B}$ (cf. Eq. (\ref{eq:matrixA}))
\begin{equation}
\widehat{A} = \widehat{B}^{-1}, \, \, B_{ij}=B_{|i-j|}=\frac{<X_{i}X_{j}>}{k_{B}T}.  
\label{eq:Bmatr_gen1}
\end{equation}
Here average is defined as the thermodynamic average with the Hamiltonian Eq. (\ref{eq:H_gen_offdiag}). 

Consider the single hole problem using the perturbation theory with respect to the electron transfer integral $b$. Then in the zeroth order approximation we set $b=0$ and assume that the hole is localized in some $G$ base numbered by index $k$. This problem can be described by the Hamiltonian 
 \begin{equation}
\widehat{H}_{int} = \frac{1}{2}\sum_{i,j=1}^{n}A_{ij}X_{i}X_{j} - X_{k}. 
\label{eq:offdiag_zeroinb}
\end{equation}
The energy minimum is realized at the point where all $n$ derivatives of energy Eq. (\ref{eq:offdiag_zeroinb}) with respect to $X_{i}$ ($i=1, .. n$) are zeros $\sum_{j=1}^{n}A_{ij}X_{j}=\delta_{ik}$. The solution to this equation is given by 
 \begin{equation}
X_{i}^{(0)}=B_{|i-k|}, 
\label{eq:eq_coord_zeroinb}
\end{equation}
and the energy minimum is  
$E_{k}^{(0)}=-\frac{B_{0}}{2}$.  

The first non-vanishing correction to the energy due to the small but finite electron transfer integral $b$ appears in the second order of perturbation theory and can be expressed as 
 \begin{equation}
\delta E_{k}^{(2)}=-\sum_{i=1}^{n}\frac{b^{2}\Delta_{ik}}{X_{i}^{(0)}-X_{k}^{(0)}},  
\label{eq:eq_energ_secondinb0}
\end{equation}
where the symbol $\Delta_{ik}$ is equal $0$ if $i$ and $k$ are non-neighboring $G$
- sites or it is equal $1$ for neighboring $G$ sites. Using the zero-point approximation for equilibrium coordinates Eq. (\ref{eq:eq_coord_zeroinb})  we get  (cf. Eq. (\ref{eq:G2offdpart}))
 \begin{equation}
\delta E_{k}^{(2)}=-z\frac{b^{2}}{B_{0}-B_{1}}=-z\frac{b^{2}}{\lambda_{*}},  
\label{eq:eq_energ_secondinb}
\end{equation}
where $z$ is the number of neighboring $G$ sites for the center of charge localization. If this site is in the edge of the $G_{n}^{+}$ sequence we get $z=1$ and if it is in the middle we get $z=2$. Thus energies of all representative states for $G_{n}^{+}$ sequences can be expressed within the second order of perturbation theory similarly to energies for the diagonal interaction (see Sec. \ref{sec:data_analysis}). The preexponential factors in all partition functions in the zeroth order in the electron transfer integral $b$ are given by factors $1/\sqrt{\det A}$, which are all the same in the top and the bottom of ratios $r_{n}$. Therefore one can still use Eq. (\ref{eq:theorGGG}) replacing the single site reorganisation energy $\lambda$ with the charge transfer reorganization energy 
\begin{equation}
\lambda_{*} = B_{0}-B_{1}.  
\label{eq:mod_reorg_en}
\end{equation}
This approximation is applicable in the regime of the strong localization. The approximate relationship between parameters of the system can be obtained using Eq. (\ref{eq:theorGGG}) with $n=3$ where $r_{n}\approx 20$. Solving this equation for the parameter in exponent we get 
\begin{equation}
\frac{b^{2}}{\lambda_{*}} \approx 0.033 {\rm eV}. 
\label{eq:b_lambd*}
\end{equation}
This is the approximate analytical relationship between the adjacent site reorganization energy and electron transfer integral. 

\subsection{Effect of adjacent $A$ bases}
\label{sec:Abases}

In the experiment \cite{FredMain} each $G_{n}$ complex is always surrounded by $A$ bases.  Ionization potential of $A$ base exceeds the one for the $G$ base by approximately $0.5$eV \cite{Voityuk1}. This potential difference is much larger than the estimated value of the electron transfer integral $b$ (see Fig. \ref{fig:potential}) so we do not expect the crucial effect due to the coupling of $G$ and $A$ bases. The coupling strength of adjacent $G$ and $A$ bases has been estimated in Ref. \cite{Voityuk}. For the equilibrium twisting angle $\theta\approx 36^{0}$ the overlap integral for the $GA$ pair is approximately the same as for 
the $GG$ pair, while for $AG$ pair it is about two times smaller (data are given for $5'-3'$ strand). If we take this effect into account using the perturbation theory approach then the ratios $r_{2}$ and $r_{3}$ take the form 
\begin{eqnarray}
r_{2} \approx \exp\left(-\frac{\beta b^2}{\lambda_{*}+\Delta} + \frac{\beta b^{2}}{\lambda_{*}}\right) + \exp\left(-\frac{\beta b^2}{4(\lambda_{*}+\Delta)} + \frac{\beta b^{2}}{\lambda_{*}}\right),
\nonumber\\
r_{3} = r_{2} + \exp\left(\frac{2\beta b^{2}}{\lambda_{*}}-\frac{\beta b^2}{\lambda_{*}+\Delta}-\frac{\beta b^2}{4(\lambda_{*}+\Delta)}\right)
\label{eq:ratA}
\end{eqnarray}
These equations can be resolved for $\lambda_{*}$ and $b$. We obtained $\lambda_{*}=0.66$eV and $b=0.18$eV. Clearly this estimate corresponds to the regime of the strong localization. The expectation for the electron transfer integral in the absence of $A$ bases for $\lambda_{*}=0.66$eV is $b=0.15$eV (see Fig. \ref{fig:potential}) so there is a  deviation between two solutions by about $20\%$.  However we cannot take this estimate quite seriously because (1) Our estimate for the electron transfer integral $b=0.18$eV exceeds the one in Ref. \cite{Voityuk} by almost the factor of $2$ (2) If we include the experimental error into our estimate then both parameters $b$ and $\lambda_{*}$ will not strongly deviate from Eq. (\ref{eq:b_lambd*}), while they may vary remarkably compared to the above estimate. Therefore the consideration of $A$ bases seems to be the excess of accuracy.  

We believe that the effect of $A$ bases can be clarified experimentally. In the sequence $TG_{n}^{+}T$ it will be more justified to ignore surrounding $T$ bases then to ignore $A$ bases in the sequence  $AG_{n}^{+}A$ studied in Ref. \cite{FredMain}. This is because the ionization potential of $T$ base exceeds that for $G$ base by the factor of $3$ more than the one for $A$ base. Therefore the measurements of ratios $r_{n}$ for $G$ sequences separated by $T$ bases will clarify the effect of neighboring bases and possibly lead to better estimates for $b$ and $\lambda$.   

One should notice that both the addition of $A$ bases and the correlations of environment coordinates  into consideration does not change our prediction that the ratios of reaction rates $r_{n}$ form the arithmetic series Eq. (\ref{eq:arithm_ser1}) in the case of the strong localization of charge.

\section{Conclusion}
\label{sec:conclusion}

We have considered the quantum state of the positive charge (hole) in poly-$G$ - poly-$C$ base sequence. We have studied a very simple but quite general model of $G_{n}$ sequence characterized by two parameters including electron  transfer integral $b$ and reorganization energy $\lambda$. Our choice of parameters $b$ and $\lambda$ was determined by the the comparison of theory and experiment for the ratios $r_{n}$ ($n=2$, $3$) of the charge transfer rate between single $G$ and $G_{n}$ complexes.   
It turns out that the agreement with the experimental data for the ratios $r_{n}$ Eq. (\ref{eq:rat1}), Ref. \cite{FredMain} can be attained only assuming the strong localization of charge within a single $G$-base.  The charge in DNA then behaves as a small polaron with the size less than the interbase distance. Charge hopping takes place due to rare environment fluctuations. Charge transfer rate is determined by the probability of such fluctuation and should obey the Arrhenius law. Based on our theory we predict that ratios of charge transfer rates form the arithmetic series $r_{n}=7.7+12.3\cdot (n-2)$. We propose the experiment measuring the ratios of rates for $n \geq 4$ as a direct verification of our theory.  However we cannot uniquely identify specific values of the electron transfer integral $b$ and the reorganization energy $\lambda$. Measurements of similar ratios in $TG_{n}T$ sequences can be used to verify the theory and to make better parameter estimates. 

In addition one can attempt to determine these parameters by measuring the temperature dependence of the charge transfer rate through poly-$G$ - poly-$C$ base sequence. We expect that this temperature dependence can be described by the Arrhenius law with the activation energy defined by the difference of charge symmetric transition state energy within $(GG)^{+}$ base sequence  Eq. (\ref{eq:pairenergyminsaddle}) and the charge ground state energy for $(GG)^{+}$ state Eq. (\ref{eq:pairenergymin}) $E_{A}=\lambda/4-b+b^{2}/\lambda$ \cite{commentnext}. Since another relationship between $b$ and $\lambda$ is known from the ratio of charge transfer rates $r_{2}$ and $r_{3}$ (see Fig.  \ref{fig:potential}) this information will be sufficient to find both parameters.  Note that the temperature range where such dependence can be measured is quite narrow ($380$K$<T<410$K) so it will be hard to prove the Arrhenius law based on such experimental data. Therefore it can be only assumption that this law is indeed applicable. Despite of the narrow temperature range the remarkable change in the charge transfer rate is expected with growing the temperature from its minimum $380$K to its maximum $410$K. For instance if $\lambda=0.3$eV this rate will increase by the factor $1.5$ with increasing temperature, while if $\lambda=1$eV then the rate will increase by the factor $3.8$.

This work is supported by the NSF CRC Program, Grant No. 0628092.
The authors acknowledge Frederick Lewis, Michael Wasielewski, George Schatz, Thorsten Fiebig, Yuri Berlin  and Mark Ratner for useful discussions.


\end{document}